\documentclass[onecolumn]{revtex4}
\usepackage{amsfonts}
\usepackage{amsmath}
\usepackage{amssymb}
\usepackage{graphicx}

\begin{document}

\title{Atomtronic superconducting quantum interference device in synthetic dimensions}
\author{Wenxi Lai}\email{wxlai@pku.edu.cn}\author{Yu-Quan Ma}\author{Yi-Wen Wei}

\affiliation{School of Applied Science, Beijing Information Science and Technology University, Beijing 100192, China}

\begin{abstract}
Coherence and scalability are essential properties of quantum systems required in quantum computers. This study presents a high coherent and scalable qubit system with atomtronics in synthetic dimensions. It is atomtronic counterpart of superconducting quantum interference device. Comparing with traditional superconducting quantum interference device which requires at least $2$-dimensional circuits, the synthetic dimensional superconducting quantum interference device can be realized only in $1$-dimensional circuits. The synthetic dimensional system is composed of Bose-Einstein condensate in two neighboring optical wells which is coupled to an external coherent light. Control parameter for the qubit is naturally provided by artificial magnetic flux originated from the coherent atom-light coupling. It should be a great advantage for the scalability and integration feature of quantum logic gates.
\end{abstract}

\maketitle

\begin{flushleft}
  \textbf{Introduction}
\end{flushleft}

Superconducting quantum interference device (SQUID) is most promising candidate of quantum system for Quantum logic gates in future quantum computers, characterized by advantages on scalability and integration~\cite{Makhlin,Fowler,Billangeon,Blais}. As well known, they are easily affected from environments due to charge and spin degree of freedom of electrons~\cite{Berkley,Simmonds,Harlingen,Faoro}. On contrast, cold atoms in cavity QED have quantum oscillations with long coherent lifetime. For example, optical oscillations in alkali earth (like) atoms enjoy lifetime about microsecond to tens of second~\cite{Ye,Livi}. The obvious drawback of cavity QED based quantum logic gates comes from their poor scalability. Although several schemes for improving the scalable feature have been proposed, they are hard to be realized and popularized in practice~\cite{Giovannetti,Duan,Lee}.

A balanced protocol may be available with atomtronics, in which atoms keep their strong coherence and, at the same time, scalability of atomic transistors becomes feasible due to similarity between atomtronic and electronic circuits~\cite{Seaman,Pepino,Lee13,Amico21}. Indeed, atomtronic circuits based on BEC Josephson junctions become one of the important candidates for the development of quantum computation. In $1$-dimensional optical potentials, BEC atoms appear quantum fluctuations of atom numbers and non local phase differences as studied both theoretically~\cite{Smerzi,Pawlowski,Laflamme,Ebgha} and experimentally~\cite{Giovanazzi,Zhu}. The Josephson junction like properties can be applied in $2$-dimensional optical potentials to realize atomtronic SQUID. They are proposed recently~\cite{Aghamalyan,Mathey,Escriva,Jezek,Obiol,Kiehn} and implemented in consequent experiments~\cite{Ryu,Ryu20,Haug}. In some literatures, such devices are directly called atomtronic quantum interference device (AQUID).

Furthermore, atomtronic flux qubits and universal quantum logic gates are demonstrated in toroidal potential wells~\cite{Solenov}, ring shaped optical lattices~\cite{Amico11,DAghamalyan} and triangular optical lattices~\cite{Safaei,Compagno}. In the ring shaped BEC, one can follow Caldeira and Leggett's dissipative model to extract effective Lagrangian of particular Josephson barriers~\cite{Solenov,Amico11,DAghamalyan}. Desired Hamiltonian for the flux qubit would be originated from the effective Lagrangian.
Couplings between two BEC rings are not hard to be implemented for the two qubit universal quantum gates as introduced in Refs. ~\cite{Aghamalyan,Escriva,Obiol,Solenov,Amico11,DAghamalyan}. Large number flux qubits are expected to be integrated in $2$-dimensional optical lattices~\cite{Safaei}.

\begin{figure}
  \includegraphics[width=8.5 cm]{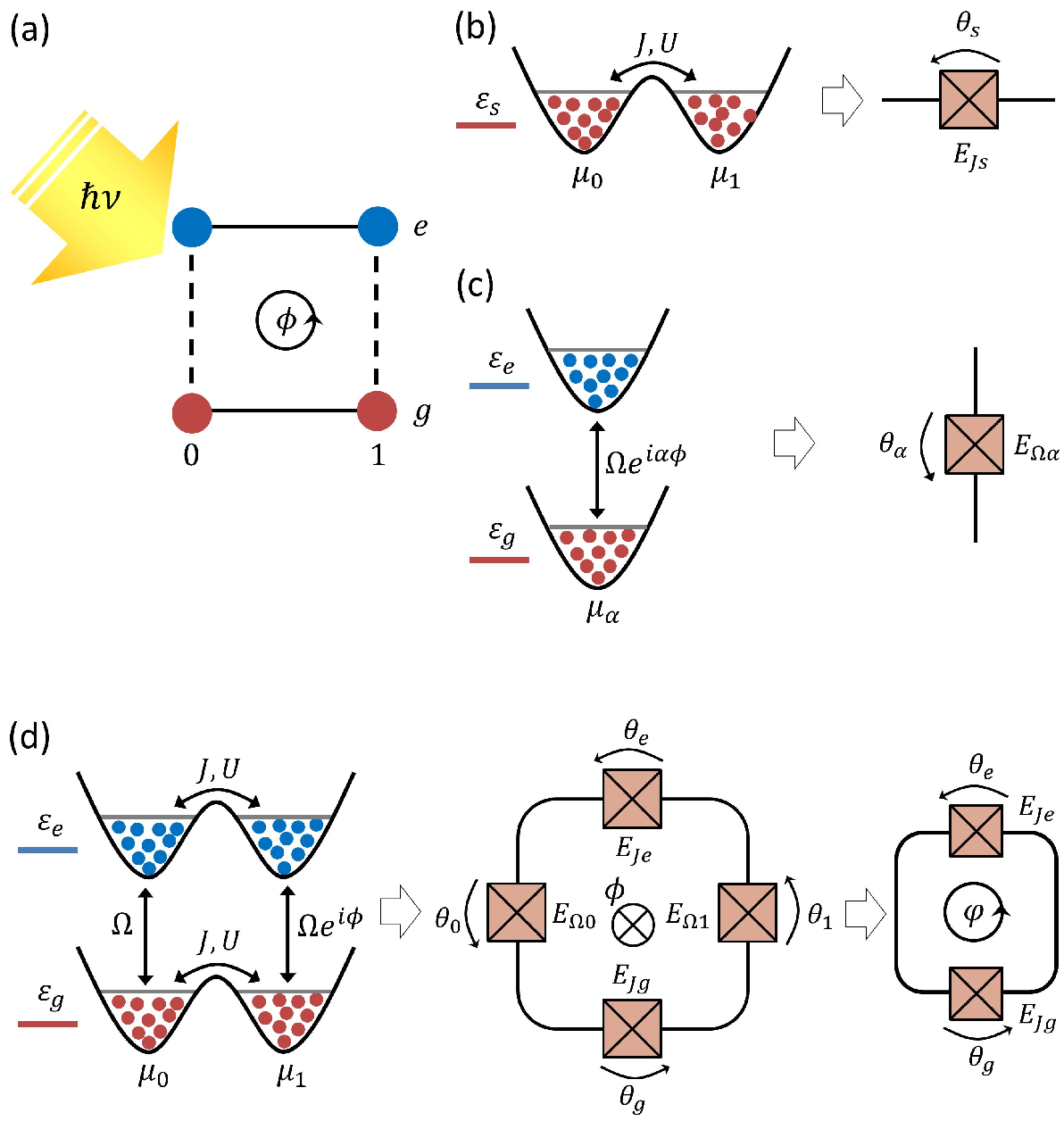}\\
  \caption{(Color on line) (a) Schematic illustration of BEC ring in synthetic dimensions. (b) Atomtronic Josephson junction in position space. (c) Atomtronic Josephson junction in atom internal states. (d) Atomtronic SQUID in synthetic dimensions.}\label{fig1}
\end{figure}

In this paper, we theoretically demonstrate atomic SQUID and flux qubit in a synthetic dimensions of BEC system. In traditional SQUID circuits, at least $2$-dimensional space is necessary to form a closed rings both in electronics~\cite{Makhlin,Fowler,Billangeon,Blais} and atomtronics~\cite{Aghamalyan,Mathey,Escriva,Jezek,Obiol,Kiehn,Ryu,Ryu20,Haug}. However, in the present synthetic dimensional SQUID, one needs only $1$-dimensional circuits, since the closed rings are created in the synthetic dimensions (Fig.~\ref{fig1} (a)). Atom-light coupling induced artificial gauge field~\cite{Livi,Celi,Wall,Huang,Mancini,Goldman} would induce magnetic flux into the synthetic dimensional ring. It could be used to control the SQUID and tunable qubits become available. Due to the lowered dimension in real space, such systems should greatly simplify the configuration of atomtronic flux qubits and improve scalability of the devices.

\begin{flushleft}
  \textbf{Atomtronic Josephson junction in position space}
\end{flushleft}

In position space, tunneling of cold atom Bose-Einstein condensate (BEC) in double-well systems can be used to realize atomic Josephson junction, which has been widely studied earlier~\cite{Smerzi,Ebgha,Zhu}. For completeness, firstly, let us review this kind of junction considering all atoms are in internal (electronic) state $s$. As shown in Fig.~\ref{fig1} (b), the typical Hamiltonian of this tunneling process is described by the Bose-Hubbard Hamiltonian $\hat{H}_{s}$~\cite{Aghamalyan,Zhu,Ryu,Haug},
\begin{eqnarray}
&&\hat{H}_{s}=\hat{H}_{Js}+\sum_{\alpha=0,1}\frac{U}{2}\hat{n}_{\alpha s}(\hat{n}_{\alpha s}-1),
\label{eq:Ham-posi}
\end{eqnarray}
where tunneling process is described by $\hat{H}_{Js}=\sum_{\alpha=0,1}\mu_{\alpha}\hat{n}_{\alpha s}-\hbar J(\hat{a}_{0s}^{\dag}\hat{a}_{1s}+H.c.)$. $\mu_{\alpha}$ represents the chemical potential of the corresponding optical well. Atom number operators are defined as $\hat{n}_{\alpha s}=\hat{a}_{\alpha s}^{\dag}\hat{a}_{\alpha s}$. Atom-atom repulsion energy due to occupying the same well is given by the quantity $\frac{U}{2}$. Rate of atom tunneling between the two wells is denoted by $J$.

Based on the mean field theory of BEC~\cite{Ryu,DAghamalyan,Orszag}, these annihilation operators could be described by the particle number and phase representation as,
\begin{eqnarray}
&&\langle \hat{a}_{\alpha s}\rangle=\sqrt{n_{\alpha s}}e^{i\theta_{\alpha s}}.
\label{eq:mean-field}
\end{eqnarray}
Then the Hamiltonian \eqref{eq:Ham-posi} becomes
\begin{eqnarray}
&&H_{s}=\sum_{\alpha=0,1}(\mu_{\alpha}n_{\alpha s}+\frac{U}{2}n_{\alpha s}^{2})-E_{Js}\cos(\theta_{1 s}-\theta_{0 s}),
\label{eq:Ham-posi-m}
\end{eqnarray}
where the Josephson energy is defined to be $E_{Js}=2\hbar J \sqrt{n_{0s}n_{1s}}$. This system has effective local capacity $C_{\alpha}=\frac{q^{2}}{U}$ which can be deduced from the comparison between the capacitance energy $\frac{Q_{\alpha s}^{2}}{2C}$ and the repulsive energy in Eq.\eqref{eq:Ham-posi-m}. At the same time, the charging energy should be $E_{C_{\alpha}}=\frac{q^{2}}{2C_{\alpha}}=\frac{U}{2}$. The effective charge is $Q_{\alpha s}=qn_{\alpha s}$. Here, $q$ is the unit charge and for neutral atoms $q=1$ with the meaning of atom number. One can use Hamilton's equations $\frac{d n_{\alpha s}}{d t}=-\frac{1}{\hbar}\frac{\partial H_{s}}{\partial \theta_{\alpha s}}$ and $\frac{d \theta_{\alpha s}}{d t}=\frac{1}{\hbar}\frac{\partial H_{s}}{\partial n_{\alpha s}}$ to reach the equations of motion,
\begin{eqnarray}
\frac{d}{dt}n_{s}=-2J \sqrt{1-n_{s}^{2}}\sin(\theta_{s})
\label{eq:mot-posi-num}
\end{eqnarray}
\begin{eqnarray}
\frac{d}{dt}\theta_{s}=\delta+\frac{U}{\hbar}N_{s} n_{s}+\frac{2Jn_{s}}{\sqrt{1-n_{s}^{2}}}\cos(\theta_{s})
\label{eq:mot-posi-pha}
\end{eqnarray}
where $\theta_{s}=\theta_{0 s}-\theta_{1 s}$, $n_{s}=\frac{n_{0 s}-n_{1 s}}{n_{0 s}+n_{1 s}}$, $N_{s}=n_{0 s}+n_{1 s}$ and $\delta=\frac{\mu_{0}-\mu_{1}}{\hbar}$. The results can be found in earlier work~\cite{Smerzi,Raghavan}. Furthermore, Eq.\eqref{eq:mot-posi-num} gives rise to the current formula in Josephson junctions~\cite{Krantz,Ryu20},
\begin{eqnarray}
I_{s}=I_{sc}\sin(\theta_{s}),
\label{eq:curr-posi}
\end{eqnarray}
where critical current of the junction is $I_{sc}=q\frac{E_{Js}}{\hbar}$.

\begin{flushleft}
  \textbf{Atomtronic Josephson junction in atom internal states}
\end{flushleft}

In the following, let us show the Josephson junction originated from coherent optical transition between two internal states ($s=g$ and $s=e$) of BEC atoms. Clock transitions of alkaline-earth (like) atoms with long lifetime can be used here. As BEC gas, all atoms are assumed to be in the same mode described with the annihilation operator $\hat{a}_{s}$, which satisfies the relation $\hat{a}_{\alpha s}|n_{\alpha s}\rangle=\sqrt{n_{\alpha s}}|n_{\alpha s}-1\rangle$. The model is shown in Fig.~\ref{fig1} (c). Optical coupling between the two internal levels of BEC atoms under a coherent light with frequency $\nu$ could be described by~\cite{Celi,Wall,Livi},
\begin{eqnarray}
&&\hat{H}_{\alpha}=\sum_{s=g,e}\varepsilon_{s}\hat{n}_{\alpha s}-\frac{\hbar \Omega}{2}(\hat{a}_{\alpha g}^{\dag}\hat{a}_{\alpha e}e^{i(\nu t-\alpha \phi)}+H.c.),
\label{eq:Ham-opt}
\end{eqnarray}
where, $\varepsilon_{s}$ denotes the ground energy level ($s=g$) or the excited energy level ($s=e$) of atom internal states. $\alpha$ represents position and $\alpha \phi$ is position dependent phase generated in coherent atom-light coupling. The Rabi frequency $\Omega$ is taken as a real quantity considering its tunable phase~\cite{Scully,Meystre}.

With the free evolution Hamiltonian $H_{\alpha 0}=\varepsilon_{g}\hat{n}_{\alpha g}+(\varepsilon_{g}+\hbar\nu)\hat{n}_{\alpha e}$, the Hamiltonian \eqref{eq:Ham-opt} is transformed into the form $H_{\alpha}^{I}$ in interaction picture.
\begin{eqnarray}
&&\hat{H}_{\alpha}^{I}=\hbar\Delta\hat{n}_{\alpha e}-\frac{\hbar \Omega}{2}(\hat{a}_{\alpha g}^{\dag}\hat{a}_{\alpha e}e^{-i\alpha \phi}+H.c.),
\label{eq:Ham-opt-int}
\end{eqnarray}

Consequently, using the mean field approach in Eq.\eqref{eq:mean-field}, $H_{\alpha}^{I}$ could be written into the $c$ number form,
\begin{eqnarray}
&&H_{\alpha}^{I}=\hbar\Delta n_{\alpha e}-E_{\Omega \alpha}\cos(\theta_{\alpha g}-\theta_{\alpha e}+\alpha \phi),
\label{eq:Ham-opt-m}
\end{eqnarray}
where the detuning $\Delta=(\varepsilon_{e}-\varepsilon_{g})/\hbar-\nu$. We define the Josephson energy, $E_{\Omega \alpha}=\hbar \Omega \sqrt{n_{\alpha g}n_{\alpha e}}$, for atomic transition between internal states under the optical coupling. The second term in Eq.\eqref{eq:Ham-opt-m} is energy of the effective inductor. Since, the model of optical transition does not involve repulsive energy like atom tunneling in the above section, the corresponding local capacity $C_{s}$ should be infinitely large, $C_{s}\rightarrow\infty$. Then, the effective charging energy is close to zero, $E_{C_{s}}=\frac{q^{2}}{2 C_{s}}\rightarrow 0$.

With Hamilton's equations $\frac{dn_{\alpha s}}{dt}=-\frac{1}{\hbar}\frac{\partial H_{\alpha I}}{\partial \theta_{\alpha s}}$, $\frac{d\theta_{\alpha s}}{dt}=\frac{1}{\hbar}\frac{\partial H_{\alpha I}}{\partial n_{\alpha s}}$, we obtain equation of motion of the flux dynamics analogous to the above method,
\begin{eqnarray}
\frac{d}{dt}n_{\alpha}=-\Omega \sqrt{1-n_{\alpha}^{2}}\sin(\theta_{\alpha}+\alpha \phi)
\label{eq:mot-opt-num}
\end{eqnarray}
\begin{eqnarray}
\frac{d}{dt}\theta_{\alpha}=-\Delta+\frac{\Omega n_{\alpha}}{\sqrt{1-n_{\alpha}^{2}}}\cos(\theta_{\alpha}+\alpha \phi)
\label{eq:mot-opt-pha}
\end{eqnarray}
where $n_{\alpha}=\frac{n_{\alpha g}-n_{\alpha e}}{n_{\alpha g}+n_{\alpha e}}$ and $\theta_{\alpha}=\theta_{\alpha g}-\theta_{\alpha e}$. Eq.\eqref{eq:mot-opt-pha} lacks the term of atom-atom repulsion energy, which is the main difference comparing with Eq.\eqref{eq:mot-posi-pha}. The repulsion energy is related to capacitance energy~\cite{Aghamalyan,Mathey}. Therefore, the optical transition of atoms just works as an effective inductor as illustrated with the equivalent circuit in Fig.~\ref{fig1} (b). Current corresponding to the phase difference can be derived from Eq.\eqref{eq:mot-opt-num} that,
\begin{eqnarray}
I_{\alpha}=I_{\alpha c}\sin(\theta_{\alpha}+\alpha \phi),
\label{eq:curr-real}
\end{eqnarray}
where critical current of the junction is $I_{\alpha c}=q\frac{E_{\Omega \alpha}}{\hbar}$.

In Eq.\eqref{eq:mot-opt-pha}, if one consider resonant coupling, $\Delta=0$ and approximately half population inversion, $n_{\alpha g}\approx n_{\alpha e}$, the phase $\theta_{\alpha}$ would be almost a constant. Recently, experiments for Josephson junction effects between two spin components of BEC atoms have been exploited with coherent optical couplings~\cite{Farolfi}. Fluctuations of spin polarization and the corresponding phase in these experiments should be originated from spin-spin interaction and the atom-light detuning which is necessary for the coherent Raman transition.

\begin{figure}
  \includegraphics[width=7.0 cm]{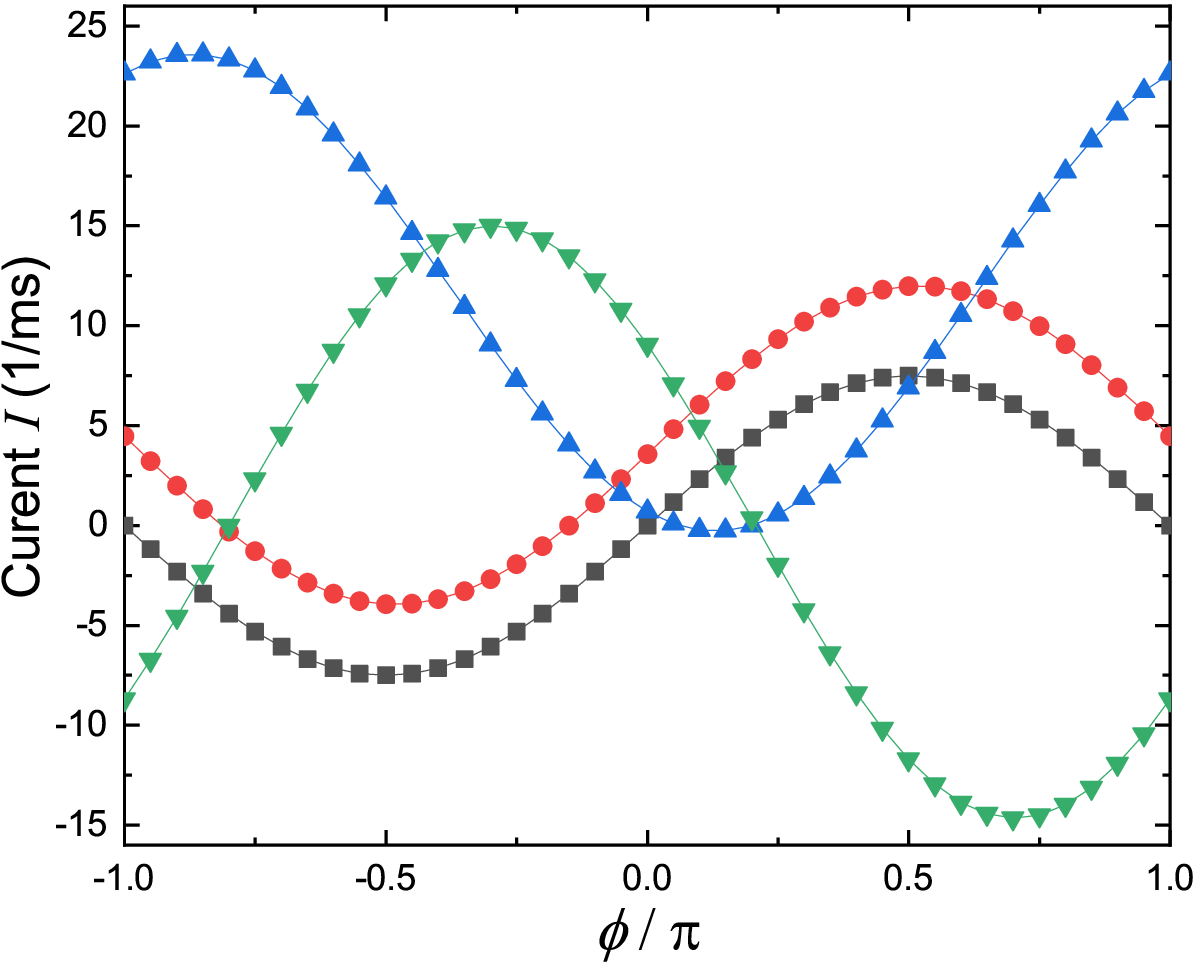}\\
  \caption{(Color on line) Atom current as a function of control phase $\phi$. Rectangular dot black line: $\theta_{0}=\theta_{1}=\theta_{g}=0$. Circle dot red line: $\theta_{0}=0.1$, $\theta_{1}=0.05$ and $\theta_{g}=0.01$. Up triangle dot blue line: $\theta_{0}=0.3$, $\theta_{1}=-0.5$ and $\theta_{g}=0.1$. Down triangle dot green line: $\theta_{0}=-0.1$, $\theta_{1}=0.9$ and $\theta_{g}=0.7$.}\label{fig2}
\end{figure}

\begin{flushleft}
  \textbf{Atomtronic SQUID in synthetic dimensions}
\end{flushleft}

When both the atom tunneling in the two-well and optical transition in atom internal states are considered at the same time, a closed loop in the synthetic dimensions would be created as illustrated in Fig.~\ref{fig1} (d). In this ring, there are four junctions all together, two junctions are in the position space and the other two are in the atom internal transition space. Hamiltonian of the ring in synthetic dimensions could be given by,
\begin{eqnarray}
&&\hat{H}_{S}=\sum_{s=g,e}\hat{H}_{Js}+\sum_{\alpha=0,1}[\frac{U}{2}\hat{N}_{\alpha}(\hat{N}_{\alpha}-1)+\hat{H}_{\alpha}],
\label{eq:Ham}
\end{eqnarray}
where, $\hat{H}_{Js}$ is given in \eqref{eq:Ham-posi}, $\hat{H}_{\alpha}$ is given by \eqref{eq:Ham-opt} and $\hat{N}_{\alpha}=\sum_{s=g,e}\hat{n}_{\alpha s}$.

Now, using the free evolution Hamiltonian $\hat{H}_{0}=\sum_{\alpha,s}(\mu_{\alpha}+\varepsilon_{s})\hat{n}_{\alpha s}$, one could transform the Eq.\eqref{eq:Ham} into its form of interaction picture,
\begin{eqnarray}
\hat{H}=\sum_{s=g,e}\hat{H}_{Js}^{I}+\sum_{\alpha=0,1}[\frac{U}{2}\hat{N}_{\alpha}(\hat{N}_{\alpha}-1)+\hat{H}_{\alpha}^{I}].
\label{eq:Ham-int}
\end{eqnarray}
where $\hat{H}_{Js}^{I}=-\hbar J(\hat{a}_{0s}^{\dag}\hat{a}_{1s}+H.c.)$. The resonant interaction $\hbar \nu=\varepsilon_{e}-\varepsilon_{g}$ and equal chemical potentials $\mu_{L}=\mu_{R}$ have been considered here.

With the mean field approach in Eq.\eqref{eq:mean-field}, the Hamiltonian \eqref{eq:Ham-int} could be written in the c number form,
\begin{eqnarray}
&&H=\sum_{\alpha=0,1}\frac{U}{2}N_{\alpha}^{2}-\sum_{s=g,e}E_{Js}\cos(\theta_{s})-\sum_{\alpha=0,1}E_{\Omega \alpha}\cos(\theta_{\alpha}+\alpha\phi).
\label{eq:Ham-int-c}
\end{eqnarray}
where $N_{\alpha}=n_{\alpha g}+n_{\alpha e}$. The linear term in the atom-atom repulsive energy is neglected due to the fact $N_{\alpha}\gg1$. Phase differences along the anti-clockwise direction have been defined as, $\theta_{g}=\theta_{0g}-\theta_{1g}$, $\theta_{1}=\theta_{1g}-\theta_{1e}$, $\theta_{e}=\theta_{1e}-\theta_{0e}$ and $\theta_{0}=\theta_{0e}-\theta_{0g}$. Algebraic sum of phase differences along the synthetic dimensional ring plus the optically induced external phase should satisfy the relation of flux quantization condition,
\begin{eqnarray}
&&\theta_{g}+\theta_{e}+\theta_{0}+\theta_{1}+\phi=2\pi m,
\label{eq:flux-quant}
\end{eqnarray}
where $m$ is an integer.

In the following, we use the Hamiltonian's equation $\frac{\partial n_{\alpha s}}{\partial t}=-\frac{1}{\hbar}\frac{\partial H}{\partial \theta_{\alpha s}}$, to derive atomic current $I$. Firstly, the Hamiltonian's equations give us equations of motion,
\begin{eqnarray}
\frac{\partial n_{\alpha g}}{\partial t}=(-1)^{\alpha}[\frac{E_{\Omega \alpha}}{\hbar}\sin(\theta_{\alpha}+\alpha\phi)-\frac{E_{Jg}}{\hbar}\sin(\theta_{g})],
\label{eq:atom-num-g}
\end{eqnarray}
\begin{eqnarray}
\frac{\partial n_{\alpha e}}{\partial t}=(-1)^{\alpha}[\frac{E_{Je}}{\hbar}\sin(\theta_{e})-\frac{E_{\Omega \alpha}}{\hbar}\sin(\theta_{\alpha}+\alpha\phi)].
\label{eq:atom-num-e}
\end{eqnarray}

Considering the anti-clockwise direction as the positive current, we have these relations for current between different islands, $q\frac{\partial n_{\alpha g}}{\partial t}=(-1)^{\alpha}(I_{\alpha}-I_{g})$ and $q\frac{\partial n_{\alpha e}}{\partial t}=(-1)^{\alpha}(I_{e}-I_{\alpha})$. The averaged formula $I=\frac{1}{4}(I_{0}+I_{g}+I_{1}+I_{e})$ gives rise to the total current in the ring,
\begin{eqnarray}
I=\frac{1}{4}[\sum_{s=g,e}I_{sc}\sin(\theta_{s})+\sum_{\alpha=0,1}I_{\alpha c}\sin(\theta_{\alpha}+\alpha\phi)].
\label{eq:current}
\end{eqnarray}
where the critical currents are $I_{sc}=\frac{q E_{J s}}{\hbar}$ and $I_{\alpha c}=\frac{q E_{\Omega \alpha}}{\hbar}$.

Due to the flux quantization condition \eqref{eq:flux-quant}, phase difference in the ring are not independent. The current formula could be obtained,
\begin{eqnarray}
I=\frac{1}{4}[I_{gc}\sin(\theta_{g})-I_{ec}\sin(\theta_{g}+\theta_{0}+\theta_{1}+\phi)+I_{0c}\sin(\theta_{0})+I_{1c}\sin(\theta_{1}+\phi)].
\label{eq:current}
\end{eqnarray}
The current formula is in accord with the current appeared in the superconducting circuit with four Josephson junctions~\cite{Qiu}. The atomtronic SQUID is in synthetic $3$-dimensional coordinates $\theta_{0}$, $\theta_{1}$ and $\theta_{g}$ with a parameter $\phi$.

Atomic currents tuned between clockwise and anti-clockwise directions are in the ring are shown in Fig.~\ref{fig2} for different phase values. Here, the tunneling rate is taken as $J=0.1$ KHz and the Rabi frequency is set to be $\Omega=0.5$ kHz which are involved in the earlier experiments~\cite{Wall,Livi}. In addition, the mean atom numbers in all islands are taken to be the same as $n_{0g}=n_{0e}=n_{1g}=n_{1e}=100$.

\begin{figure}
  \includegraphics[width=12.0 cm]{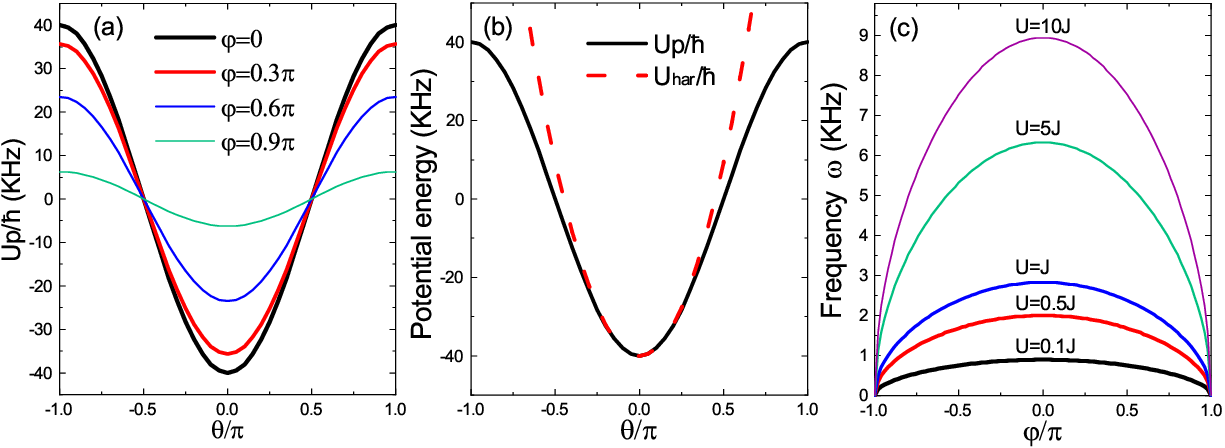}\\
  \caption{(Color on line) (a) Potential energy versus phase $\theta$. Here, $U/\hbar=10J$. (b) Comparison between the original potential and quadratic order potential with respect to $U/\hbar=10J$. (c) Frequency of the harmonic oscillator defined in Eq.\eqref{eq:Ham-quant}.}\label{fig3}
\end{figure}

\begin{flushleft}
  \textbf{Atomtronic flux qubit in synthetic dimensions}
\end{flushleft}

Model for the Hamiltonian \eqref{eq:Ham-int-c} is shown in Fig.~\ref{fig1} (d). During the resonant coupling of coherent optical transition, we have $\Delta=0$ and $n_{\alpha g}\approx n_{\alpha e}$. From Eq. \eqref{eq:mot-opt-pha} one can conclude that corresponding phase fluctuation could be negligible small, namely, $\dot{\theta}_{\alpha}=\dot{\theta}_{\alpha g}-\dot{\theta}_{\alpha e}\rightarrow 0$. The negligible small phase change is reasonable since atom-atom repulsion energy is not involved when atoms transit between its two internal states $g$ and $e$. Therefore, we take phase differences $\theta_{0}$ and $\theta_{1}$ as time independent classical quantities. Comparing Eqs.\eqref{eq:mot-posi-pha} and \eqref{eq:mot-opt-pha}, it is reasonable to assume that phase fluctuation in the loop is mainly contributed from the atom tunneling between the two wells, namely, $\dot{\theta}_{s}=\dot{\theta}_{Ls}-\dot{\theta}_{Rs}\neq 0$. Under this condition, Hamiltonian \eqref{eq:Ham-int-c} is reduced into,
\begin{eqnarray}
&&H=\sum_{\alpha=0,1}\frac{U}{2}N_{\alpha}^{2}-\sum_{s=g,e}E_{Js}\cos(\theta_{s}).
\label{eq:Ham-int-c2}
\end{eqnarray}
where the non-dynamical constant terms $-E_{\Omega 0}\cos(\theta_{0})$ and $-E_{\Omega 1}\cos(\theta_{1}+\phi)$ have been neglected since the phases $\theta_{0}$ and $\theta_{1}$ are regarded as time independent classical parameters.

The Hamiltonian \eqref{eq:Ham-int-c2} is expressed in terms of general coordinates and momentums, $H=E(n_{\alpha s}, \theta_{\alpha s})$. To describe the Hamiltonian in terms of velocities and coordinates, $H=E(\dot{\theta}_{\alpha s}, \theta_{\alpha s})$, one may find the corresponding velocities from the Hamiltonian's equation, $\dot{\theta}_{\alpha g}=\frac{1}{\hbar}\frac{\partial H}{\partial n_{\alpha g}}$~\cite{Tsvelik}. It gives rise to the results
\begin{eqnarray}
\frac{d\theta_{0s}}{dt}=\frac{U}{\hbar}(n_{0 g}+n_{0 e})-J\sqrt{\frac{n_{1s}}{n_{0s}}}\cos(\theta_{s}),
\label{eq:phas-0s}
\end{eqnarray}
\begin{eqnarray}
\frac{d\theta_{1s}}{dt}=\frac{U}{\hbar}(n_{1 g}+n_{1 e})-J\sqrt{\frac{n_{0s}}{n_{1s}}}\cos(\theta_{s}).
\label{eq:phas-1s}
\end{eqnarray}
When the quantity $\frac{U}{\hbar}$ is comparable to $J$, and $n_{0s}$, $n_{1s}$ $\gg1$, this inequality $\frac{U}{\hbar}(n_{\alpha g}+n_{\alpha e})\gg J\sqrt{\frac{n_{0s}}{1s}}$ should be satisfied. In this regime, the second term in Eqs. \eqref{eq:phas-0s} and \eqref{eq:phas-0s} could be neglected. Then, we obtain the relations $\dot{\theta}_{0s}\approx\frac{U}{\hbar}(n_{0 g}+n_{0 e})$ and $\dot{\theta}_{1s}\approx\frac{U}{\hbar}(n_{1 g}+n_{1 e})$ for $s=g$, $e$. These two relations, allow us to write $(n_{\alpha g}+n_{\alpha e})\approx\frac{\hbar}{2U}(\dot{\theta}_{\alpha g}+\dot{\theta}_{\alpha e})$ for $\alpha=0$, $1$. One can take $N_{\alpha}$ into the Hamiltonian \eqref{eq:Ham-int-c2},
\begin{eqnarray}
&&H=\sum_{\alpha=0,1}\frac{\hbar^{2}}{8U}(\dot{\theta}_{\alpha s}+\dot{\theta}_{\alpha e})^{2}-\sum_{s=g,e}E_{Js}\cos(\theta_{s}).
\label{eq:Ham-int-c3}
\end{eqnarray}
Considering the phase differences $\theta_{g}=\theta_{0g}-\theta_{1g}$ and $\theta_{e}=\theta_{1e}-\theta_{0e}$, it is convenient to define $\theta_{0g}=\Theta_{g}+\frac{\theta_{g}}{2}$, $\theta_{1g}=\Theta_{g}-\frac{\theta_{g}}{2}$, $\theta_{0e}=\Theta_{e}-\frac{\theta_{e}}{2}$ and $\theta_{1e}=\Theta_{e}+\frac{\theta_{e}}{2}$, where $\Theta_{g}$ and $\Theta_{e}$ are time independent mean quantities. They leads to the Hamiltonian in terms of non local phases,
\begin{eqnarray}
&&H=\frac{\hbar^{2}}{16U}(\dot{\theta}_{g}-\dot{\theta}_{e})^{2}-\sum_{s=g,e}E_{Js}\cos(\theta_{s}).
\label{eq:Ham-int-c4}
\end{eqnarray}
It reflects the synthetic dimensional atomtronic SQUID is equivalent to the SQUID containing two Josephson junctions on a ring~\cite{Rasmussen}. Schematic illustration of this configuration is shown at the bottom of Fig.~\ref{fig1} (d).

Because $\theta_{0}$ and $\theta_{1}$ are time independent classical quantities, we absorb $\theta_{0}$ and $\theta_{1}$ into $\phi$ and define $\varphi=\theta_{0}+\theta_{1}+\phi$. In this case, the flux quantized condition \eqref{eq:flux-quant} becomes $\theta_{g}+\theta_{e}+\varphi=2m\pi$. It reveals there is just one degree of freedom in this system except $\varphi$. Then, Eq. \eqref{eq:Ham-int-c4} would be reduced into the following simple form,
\begin{eqnarray}
&&H=\frac{\hbar^{2}}{4U}\dot{\theta}_{g}^{2}-E_{Jg}\cos(\theta_{g})-E_{Je}\cos(\theta_{g}+\varphi).
\label{eq:Ham-int-c5}
\end{eqnarray}
Now, one may choose a new origin of coordinate that let $\theta_{g}=\theta-\frac{\varphi}{2}$. Additionally, considering $E_{Jg}\approx E_{Je}$ in the resonant coherent coupling, one can find,
\begin{eqnarray}
&&H_{eff}=\frac{\hbar^{2}}{4U}\dot{\theta}^{2}-2E_{Jg}\cos(\frac{\varphi}{2})\cos(\theta).
\label{eq:Ham-int-v6}
\end{eqnarray}
This is the Hamiltonian of single qubit with tunable parameter $\varphi$~\cite{Krantz}. It is valuable to point out that, definition of the phase difference $\dot{\theta}=\dot{\theta}_{Lg}-\dot{\theta}_{Rg}$ and $\dot{\theta}_{\alpha g}\approx\frac{U}{\hbar}(n_{\alpha g}+n_{\alpha e})$ given above would actually result in the relation between general velocity and momentum, $\dot{\theta}=\frac{U}{\hbar}Nn$. Here, $N$ is the total atom number $N=N_{0}+N_{1}$ and $n$ is the atom number difference $n=(N_{0}-N_{1})/N$ between two wells.

From Faraday's law, one know that the effective voltage $V$ should equal to $V=\frac{\hbar}{q}\dot{\theta}$. On the other hand, the effective capacitance $C$ satisfies $C=\frac{Q}{V}$, where $Q=qn$ and $n$ represents atom number with respect to the phase $\theta$. Then, one find the relation $Q=\frac{C \hbar}{q}\dot{\theta}$. It allow us to obtain energy of the effective capacitance $\frac{Q^{2}}{2C}=\frac{C \hbar^{2}}{2 q^{2}}\dot{\theta}^{2}$. Comparing with the capacitance energy $\frac{Q^{2}}{2C}$ and the kinetic energy in Hamiltonian \eqref{eq:Ham-int-v6}, one find the capacitance equals to $C=\frac{q^{2}}{2U}$. Finally, from the charging energy formula $E_{C}=\frac{q^{2}}{2C}$, we have $E_{C}=U$. When the charging energy satisfies $E_{C}\gg E_{Jg}$, phase fluctuation in the junction is large, on the contrary, when the charging energy satisfies $E_{C}\ll E_{Jg}$, phase fluctuation is small~\cite{Rastelli}.

Fig.~\ref{fig3} (a) illustrates that the external control phase could arbitrarily change depth of the potential $U_{P}=-2E_{Jg}\cos(\frac{\varphi}{2})\cos(\theta)$. Anharmonic property of this potential allows us to access a uniquely addressable quantum two-level system for qubit designing~\cite{Krantz}. From Taylor expansion of the term $\cos(\theta)$,
\begin{eqnarray}
&&H_{eff}=\frac{\hbar^{2}}{4U}\dot{\theta}^{2}-2E_{Jg}\cos(\frac{\varphi}{2})(1-\frac{\theta^{2}}{2}+\frac{\theta^{4}}{24}+...).
\label{eq:Ham-int-qua}
\end{eqnarray}
one can extract quadratic order of the potential $U_{har}=-2E_{Jg}\cos(\frac{\varphi}{2})(1-\frac{\theta^{2}}{2})$. Fig.~\ref{fig3} (b) shows comparison between the effective potential $U_{P}$ and its quadratic part $U_{har}$. For small $\theta$, we have the approximation $U_{P}\rightarrow U_{Q}$. Frequency $\omega$ of the harmonic oscillator $H_{har}=\frac{\hbar^{2}}{4U}\dot{\theta}^{2}+U_{Q}$ is not hard to be given by,
\begin{eqnarray}
&&\omega=\frac{2}{\hbar}\sqrt{UE_{Jg}\cos(\frac{\varphi}{2})}.
\label{eq:leading-fre}
\end{eqnarray}
Here, $\varphi$ should takes $2m\pi-\pi<\varphi<2m\pi+\pi$. Just considering lowest two levels of the Hamiltonian $H_{har}$, one could write an estimated Hamiltonian of two-level system,
\begin{eqnarray}
&&\hat{H}_{Q}= \hbar \omega \sigma_{z},
\label{eq:Ham-quant}
\end{eqnarray}
where $\sigma_{z}$ is the Pauli matrix. Eq. \eqref{eq:Ham-int-v6} represents the free evolution Hamiltonian of the qubit system whose energy level could be tuned by the system parameters as illustrated in Fig.~\ref{fig3} (c). All dates in this figure belong to the regime $E_{C}\ll E_{Jg}$. The condition $E_{C}\gg E_{Jg}$ should occur when the total atom number is very small and optical potentials are very narrow, so as to cause atom-atom repulsion is much intensive.

\begin{flushleft}
  \textbf{Conclusions}
\end{flushleft}

In conclusions, quantum tunneling of BEC between two neighboring wells and optical transitions in the atom internals states could form SQUID in synthetic dimensions. Clockwise and anti-clockwise currents in the synthetic dimensional ring can be predicted. Optical transitions in atom-light interactions work as effective inductors and appear the property of Josephson junctions. In the optical Josephson junction, atom number fluctuation is negligible during the coherent resonant coupling. This feature would be favorable to simplify four-junction SQUID into two-junction SQUID. Consequently, considering the relation of flux quantization in the closed loop, one could achieve $1$-dimensional quantum phase fluctuation in the atomtronic SQUID. Therefore, the model may have great applications for the qubit design in quantum computations. This qubit is controllable with the artificial magnetic flux generated from the coherent atom-light coupling.

\begin{acknowledgments}
This work was supported by National Natural Science Foundation of China (Grant No. 12304190), Natural Science Foundation of Beijing Municipality (Grant No. 1252018), R \& D Program of Beijing Municipal Commission of Education (Grant No. KM202011232017), and Natural Science Foundation of Beijing Municipality (Grant No. 1232026).
\end{acknowledgments}

\emph{Data availability statement}: All data that support the findings of this study are included within the article.

\end{document}